\documentclass[10pt,letterpaper]{article}
\usepackage{opex3}
\usepackage{subeqn}
\usepackage{url}
\usepackage[normalem]{ulem}
\usepackage{cite}
\usepackage[]{mcode}
\input{My_def_Opex.def}
\begin{document}

\title{Accumulation of nonlinear interference noise in fiber-optic systems}

\author{Ronen Dar,$^1$ Meir Feder,$^1$ Antonio Mecozzi,$^2$ and Mark Shtaif$^{1,*}$}
\address{$^1$School of Electrical Engineering, Tel Aviv University, Tel Aviv, Israel 69978\\$^2$Department of Physical and Chemical Sciences, University of L'Aquila, 67100 L'Aquila, Italy}
\email{* shtaif@eng.tau.ac.il }

\begin{abstract*}
Through a series of extensive system simulations we show that all of the previously not understood discrepancies between the Gaussian noise (GN) model and simulations can be attributed to the omission of an important, recently reported, fourth-order noise (FON) term, that accounts for the statistical dependencies within the spectrum of the interfering channel. {We examine the importance of the FON term as well as the dependence of NLIN on modulation format with respect to link-length and number of spans}. A computationally efficient method for evaluating the FON contribution, as well as the overall NLIN power is provided.
\end{abstract*}

\vspace{1cm}

\section{Introduction}
Inter-channel nonlinear interference is arguably the most important factor in limiting the performance of fiber-optic communications \cite{Essiambre}. Since joint processing of the entire WDM spectrum of channels is prohibitively complex, nonlinear interference between channels is customarily treated as noise. The statistical characterization of this noise --- to which we refer in what follows as nonlinear interference noise (NLIN) --- is the goal of most recent theoretical studies of nonlinear transmission \cite{Poggiolini,Carena,PogECOC,Johannisson,DarOpex,Mecozzi,Secondini,PogArxiv}. Understanding the features of NLIN is critical for the efficient design of fiber-optic systems and for the accurate prediction of their performance.

Most of the available work on NLIN in fiber-optic systems was published in the context of the Gaussian noise (GN) model \cite{Poggiolini,Carena,PogECOC,Johannisson}, which describes NLIN as an additive Gaussian noise process whose variance and spectrum it evaluates. The validation of the GN model and the characterization of its accuracy have been the subject of numerous studies (e.g. see \cite{Carena,PogECOC}). It was found that while the model's accuracy is satisfactory in some scenarios, it is highly inadequate in others. Some of the GN model's most conspicuous shortcomings are its independence of modulation format \cite{DarOpex}, its independence of pre-dispersion \cite{PogECOC}, and its large inaccuracy in predicting the growth of the NLIN variance with the number of spans in an amplified multi-span link \cite{Carena,PogECOC}. Although phenomenological fixes for the latter problem have been proposed (most notably through the practice of accumulating the NLIN contributions of various spans incoherently \cite{Poggiolini,Carena,PogECOC}), the remedy that they offered remained limited, and the fundamental reason for the observed behavior has never been understood.

We argue, similarly to \cite{DarOpex}, that the reason for the inaccuracy of the GN approach is in ignoring the statistical dependence between different frequency components of the interfering channel. Accounting for this dependence produces an important correction term to which we refer (for reasons explained in the following section) as the \emph{fourth-order noise} or FON. By simulating a number of fiber-systems in the relevant range of parameters, we demonstrate that the FON term resolves all of the reported inaccuracies of the GN model, including the dependence on modulation format, signal pre-dispersion, and the accumulation of NLIN with the number of spans.

We stress that as demonstrated in \cite{DarOpex} the NLIN is not an additive Gaussian process, and hence its variance (and even its entire spectrum) does not characterize its properties in a satisfactory manner. For example, as pointed out in \cite{Mecozzi}, part of the NLIN manifests itself as phase-noise, whose effect in terms of transmission performance is very different from that of additive noise \cite{DarOL}. We show in what follows that the phase-noise character of NLIN, as well as the dependence of NLIN on modulation format is largest in the case of a single amplified span, or in a system of arbitrary length that uses distributed amplification (as in \cite{DarOpex}). The distinctness of these properties reduces somewhat in multi-span systems with lumped amplification and with a span-length much larger than the fiber's effective length.

In order to facilitate future research of this problem, we provide a computer program that implements a computationally efficient algorithm for computing the SON and the FON coefficients that are needed for reproducing the theoretical curves that we present in this paper. {For the reader's convenience,  the program also includes the option of computing the entire NLIN variance, including intra-channel interference terms, as well as inter-channel interference terms that are not directly addressed in the main text of this paper, and which have been recently posted in}  \cite{PogArxiv}. {The contribution of these terms reduces very rapidly with channel spacing and while one may generate situations in which inclusion of these terms becomes relevant, they were negligible in the system studied in} \cite{DarOpex} {(which assumed distributed amplification, and a guard-band as small as 2\%), and they are also negligible in systems with more realistic parameters, as considered here.} We note that while the full scale simulations of the systems of interest are computationally intense and time consuming, the extraction of NLIN power on the basis of the FON and SON coefficients that we provide is practically instantaneous.

\section{Theoretical background}
In a recent paper \cite{DarOpex} we have demonstrated that by removing the assumption of statistical independence between frequency components within the interfering channel (which has been used in the derivation of the GN model) the variance of NLIN is given by
\bea \sigma_{\mathrm{NLIN}}^2 = \underbrace{\vphantom{P^3\chi_2\left(\frac{\lip |b|^4\rip}{\langle |b|^2\rangle^2} -2\right)}P^3\chi_1}_{\mathrm{SON\setminus GN}} + \underbrace{P^3\chi_2\left(\frac{\lip |b|^4\rip}{\langle |b|^2\rangle^2} -2\right)}_{\mathrm{FON}},\label{10}\eea
where $P$ is the average power, $b$ denotes the data symbol in the interfering channel (e.g. for QPSK modulation $b$ is a random variable that receives each of the four values $\pm\frac{1}{\sqrt{2}}\pm \frac{i}{\sqrt{2}}$ with probability of 1/4), and the angled brackets denote statistical averaging. The terms $\chi_1$ and $\chi_2$ are given by Eqs. (26--27) in \cite{DarOpex} multiplied by $T^3$, where $T$ is the symbol duration. These coefficients are functions of the transmitted pulse waveform and of the fiber parameters. The first term on the right-hand-side of (\ref{10}) is identical to the result of the GN model, and since it follows only from second-order statistics, we refer to it as the second-order noise (SON) term. The second term depends on fourth-order statistics and is hence referred to as the fourth-order noise (FON) term. The presence of $\langle |b|^4\rangle$ in the FON term implies modulation format dependence. For example, the NLIN variance is $ P^3\left(\chi_1 -\chi_2\right)$ with QPSK modulation, $P^3\left(\chi_1 -0.68\chi_2\right)$ with 16-QAM, and $P^3\chi_1$ when Gaussian modulation is used. Note that only with Gaussian modulation the NLIN variance is independent of $\chi_2$ and hence the GN-model's prediction is exact. In the section that follows we demonstrate the accuracy of Eq. (\ref{10}) with respect to a range of fiber-optic systems that we simulate, and discuss the role and relative importance of the FON term in the various scenarios.

We note that in order to compare with the theory of \cite{Poggiolini,Carena,PogECOC,Johannisson}, the SON and FON coefficients were written in Eqs. (26) and (27) of \cite{DarOpex} without including the band-limiting effect of the receiver matched filter. To include its effect, products that fall outside of the received channel bandwidth should be excluded from the summation. The computer program which we provide in the appendix to this paper for the extraction of $\chi_1$ and $\chi_2$ {accounts for the presence of a matched filter}.

\section{Results}
The results are obtained from a series of simulations considering a five-channel WDM system implemented over standard single-mode fiber (dispersion of 17 ps/nm/km, nonlinear coefficient $\gamma = 1.3$ [Wkm]$^{-1}$, and attenuation of 0.2dB per km). We assume Nyquist pulses with a perfectly square spectrum, a symbol-rate of 32 {GSymbols/s} and a channel spacing of 50 GHz. {The number of simulated symbols in each run was 4096 and the total number of runs that were performed with each set of system parameters (each with independent and random data symbols)  ranged between 100 and 500 so as to accumulate sufficient statistics}. As we are only interested in characterizing the NLIN, we did not include amplified spontaneous emission (ASE) noise in any of the simulations. At the receiver, the channel of interest was isolated with a matched optical filter and {ideally} back-propagated so as to eliminate the effects of self-phase-modulation and chromatic dispersion. All simulations were performed with a single polarization, whereas the scaling of the theoretical results to the dual polarization case has been discussed in \cite{DarOpex}. {For both forward and backward propagation, the scalar nonlinear Schr\"{o}dinger equation has been solved using the standard split-step-Fourier method with a step size that was set to limit the maximum nonlinear phase variation to 0.02 degrees (and bounded from above by 1000 m). The sampling rate was 16 samples per-symbol. To extract the NLIN, we first removed the average phase-rotation induced by the neighboring WDM channels and then evaluated the offset between the received constellation points and the ideal constellation points that would characterize detection in the absence of nonlinearity.}

In Fig. 1 we show the NLIN power as a function of the average input power for a system consisting of $5\times 100$ km spans in the cases of QPSK and 16-QAM modulation. Figure 1(a) corresponds to the case of purely distributed amplification whereas Fig. 1(b) represents the same system in the case of lumped amplification. The solid curves represent the analytical results obtained from Eq. (\ref{10}) while the dots represent the results of the simulations. The dashed red curve shows the prediction of the GN model, i.e. $P^3\chi_1$. The dependence on modulation format is evident in both figures, as is the GN model's offset. However, while the error of the GN model in the case of QPSK is 10dB for distributed amplification, it reduces to 3.7dB when lumped amplification is used. Note that the difference between the modulation formats, as well as the error of the GN-model result are both independent of the input power. The excellent agreement between the theory (Eq. (\ref{10})) and simulation is self evident.

\begin{figure}
\begin{centering}
\includegraphics[width=0.75\columnwidth]{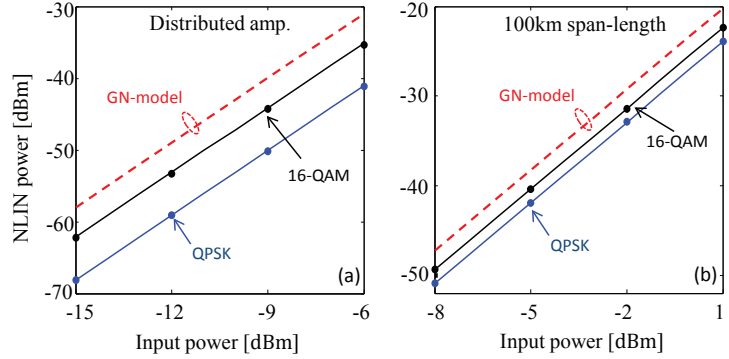}
\caption{The NLIN power versus the average power per-channel in a $5\times 100$km system for QPSK and 16-QAM modulation. The solid lines show the theoretical results given by Eq. (\ref{10}) and the dots represent simulations. The dashed red line corresponds to the SON contribution $P^3\chi_1$, which is identical to the result of the GN model. (a) Distributed amplification. (b) Lumped amplification.}\label{NLINvsP}
\end{centering}
\end{figure}

More insight as to the significance of the span-length can be extracted from Fig. 2 which shows the received constellations in a 500 km system for QPSK (top panels) and 16-QAM (bottom panels) transmission. The first column of panels from the left [Figs. 2(a) and 2(b)] correspond to the case of distributed amplification, the second column of panels [Figs. 2(c) and 2(d)] correspond to 25 km spans, the third column of panels [Figs. 2(e) and 2(f)] correspond to 50 km spans and the case of 100 km spans is shown in the rightmost panels [Figs. 2(g) and 2(h)]. Here and in the figures that follow, {the launched} powers were selected such that the path-averaged power {per-channel} was -10dBm in all cases (input power of -7.7dBm, -5.9dBm and -3.3dBm, for 25 km, 50 km, and 100 km spans, respectively). Use of a constant path-averaged power is customary when comparing systems with different span lengths \cite{GordonMollenauer}, and the value of -10dBm was found to be roughly optimal from the standpoint of capacity maximization in the distributed amplification case \cite{Mecozzi}.
As can be seen in Fig. 1, the comparison between modulation formats is practically independent of the launched optical power.
Consistently with the predictions in \cite{DarOpex}, the phase noise is negligible in the case of QPSK transmission, and the constellation points are nearly circular for all span-lengths. With 16-QAM modulation, phase-noise is dominant in the case of distributed amplification and 25 km spans, whereas at longer span-lengths the phase-noise character of NLIN becomes less evident. We note that even in the case of 100 km spans, phase noise is clearly visible implying that NLIN cannot be accurately described as additive, although the error from doing so is notably smaller.

\begin{figure}
\begin{centering}
\includegraphics[width=1.0\columnwidth]{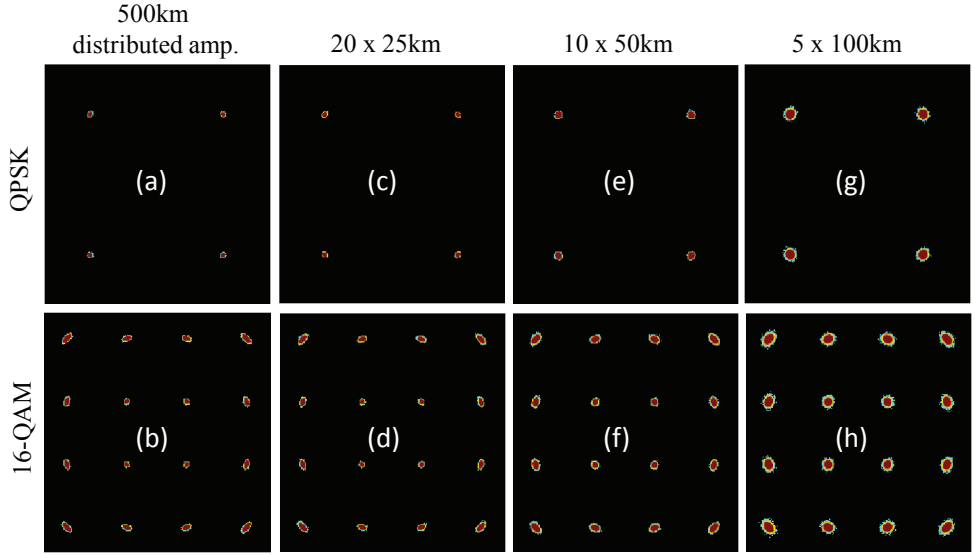}
\caption{Constellation diagrams for QPSK (top panels) and 16-QAM (bottom panels) in the cases of distributed amplification (a and b), 25 km spans (c and d), 50 km (e and f) and 100 km (g and h). In all cases a path-averaged power of -10dBm was used. The phase noise nature of NLIN is evident in the case of 16-QAM modulation, but its relative significance reduces when the span-length is large. The 100 km span case is closer to the circular noise distributions observed in \cite{Carena}.}\label{Const}
\end{centering}
\end{figure}

Figure 3 shows the accumulation of NLIN along systems of different span-lengths. Fig. 3(a) shows the case of distributed amplification whereas Figs. 3(b)-3(d) show the cases of 25 km, 50 km, and 100 km span-lengths, respectively. The NLIN power in these figures is normalized to the received optical power in each case so that the vertical axes can be interpreted as noise to signal ratio. Notice that in the cases of single-span transmission the dependence on modulation format and the inaccuracy of the GN model are largest. As the number of spans increases the NLIN increases more rapidly for the QPSK format and the difference between the modulation formats reduces. After 500 km of propagation the difference between the NLIN power for QPSK and 16-QAM modulation formats is approximately 6dB, 4.8dB, 2.8dB and 1.5dB for distributed amplification, 25 km, 50 km, and 100 km spans. The error of the GN model in the case of QPSK is approximately 10dB, 8.6dB, 5.8dB, and 3.7dB for the cases of distributed amplification, 25 km, 50 km and 100 km spans.

\begin{figure}
\begin{centering}
\includegraphics[width=0.75\columnwidth]{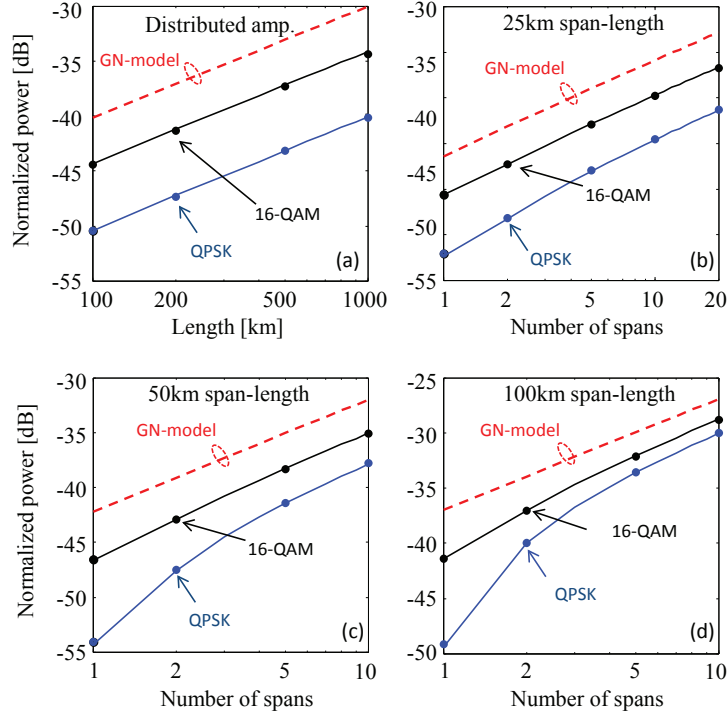}
\caption{Accumulation of the NLIN power (normalized to the received power) with the number of spans. Figure a corresponds to the case of distributed amplification whereas figures b,c and d correspond to the cases of 25 km, 50 km, and 100 km span-lengths, respectively. The solid lines show the theoretical results given by Eq. (\ref{10}) and the dots represent simulations. The red dashed curve represents the SON, or equivalently the GN model result.}\label{Accum}
\end{centering}
\end{figure}

\section{Discussion}
The explanation to the observed behavior can be attributed to the dynamics of nonlinear collisions in WDM systems, whose details will be discussed in a separate publication. In particular, NLIN that is induced by cross-phase-modulation (XPM) is strongest when the nonlinearly interacting pulses experience incomplete collisions \cite{Shtaif98}. In a single-span system, or in a system with distributed gain, such collisions occur only twice; once in the beginning of the system and again at its end. In a multi-span system with lumped amplification incomplete collisions occur at every point of power discontinuity, namely at the beginning of every amplified span. Incomplete collisions taking place at different locations, produce NLIN contributions of independent phase and therefore, when the NLIN is dominated by incomplete collisions, it appears more isotropic in phase-space and its distribution becomes closer to Gaussian. {The relative significance of incomplete collisions is determined mainly by two factors; the number of incomplete collisions (which grows with the number of spans), and the magnitude of power discontinuity (increases with span length). For a fixed length link, when the number of spans is so large that attenuation within the span is negligible, the power discontinuity at the amplifier sites vanishes and the system becomes equivalent to a distributed gain system, where only two incomplete collisions occur (at the beginning and at the end of the entire link). As the span length increases, the power discontinuity at the amplifier locations grows and the overall significance of incomplete collisions increases.
This explains the fact that in Fig. \ref{Const} the 16-QAM constellation spots appear more and more circular as the span length increases from 0 (distributed amplification) to 100 km. When the span-length increases further, to the extent that it becomes much longer than the fiber's effective length ($1/\alpha \sim 20$ km in most fibers), the growth in the power discontinuity becomes negligible, but the number of incomplete collisions continues to decrease with the number of spans, until eventually, in a single-span link (where only one incomplete collision occurs at the link's beginning) the non-Gaussianity of NLIN reappears and the deviation from the GN model is very significant.
This point can be seen in Figs. 3(b)--3(d), where the NLIN variance is shown as a function of the number of spans and the span-length is kept constant. The error in the GN model is always largest in a single-span link, and reduces considerably with the number of spans in the case of lumped amplification. }

\begin{figure}
\begin{centering}
\includegraphics[width=0.75\columnwidth]{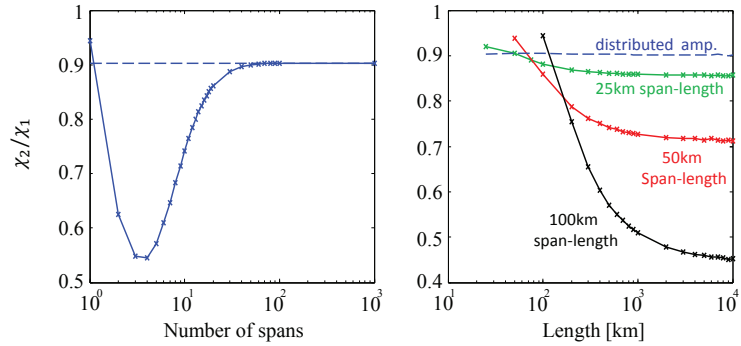}
\caption{The importance of incomplete collisions is reflected in the ratio between the FON and SON coefficients $\chi_2/\chi_1$. When incomplete collisions dominate $\chi_2/\chi_1\ll 1$, and when their contribution is small (as occurs in single span, or distributed gain systems) $\chi_2/\chi_1\sim 1$. In (a) The total link-length is held fixed at 500 km. In (b) the span length is kept constant. In both Figs. (a) and (b), the dashed curve corresponds to distributed amplification.}\label{chi2chi1}
\end{centering}
\end{figure}

{Figure \ref{chi2chi1} summarizes these ideas by showing the ratio $\chi_2/\chi_1$ as a function of the number of spans in a fixed-length system [Fig. \ref{chi2chi1}(a)], and as a function of system length [Fig. \ref{chi2chi1}(b)]. When incomplete collisions dominate so that the NLIN approaches a circular Gaussian distribution and the significance of phase-noise reduces, $\chi_2/\chi_1\ll 1$ and the NLIN variance is dominated by the SON contribution (the GN model). In Fig. \ref{chi2chi1}(a) the deviation from the GN model is seen to be largest ($\chi_2/\chi_1\sim 1$) in the single-span case, and when the number of spans is so large that the scheme approaches the conditions of distributed amplification. When the span-length is fixed, as in Fig. \ref{chi2chi1}(b), the ratio $\chi_2/\chi_1$ reduces with the length of the link, with the highest rate of reduction occurring when the span-length is long so that the power discontinuity is largest.}

Another interesting aspect of the nonlinear dynamics is revealed in the context of the effect of signal pre-dispersion. One of the most central claims made in \cite{Poggiolini,Carena,PogECOC,PogArxiv} is that signal Gaussianity, which is crucial for the validity of the GN model, follows from the accumulated effect of chromatic dispersion, and hence the large inaccuracy of the GN model in the first few spans of a WDM system was attributed to the fact that the signal is not sufficiently dispersed. Indeed, in \cite{PogECOC} it has been demonstrated that in the presence of very aggressive pre-dispersion, the NLIN variance is accurately described by the GN model even in the very first few spans {(where without pre-dispersion the inaccuracy of the GN model is largest)}. In our understanding the role of dispersion in this context has been misconstrued. {While it is true that significant pre-dispersion reduces the GN model's inaccuracy in the first span, as shown in} \cite{PogECOC}, {it is not the absence of sufficient dispersion that explains the GN model's inaccuracy.} 
{Here we present an alternative view at the role of pre-dispersion.} We plot in Fig. \ref{Accum} the NLIN variance as a function of system length, once in the case of distributed amplification and once in the case of lumped amplification with 100km spans. In both cases the signals were pre-dispersed by 8500 ps/nm --- equivalent to a 500 km long link. Notice that indeed pre-dispersion improves the accuracy of the SON term representing the GN model in the first few spans. However, when the link becomes longer and the accumulated dispersion exceeds the amount of pre-dispersion assigned to the signal, the deviation from the GN-model increases and eventually, the simulated NLIN variance approaches the same value that it has without pre-dispersion. This behavior is seen to be in clear contrast to the interpretation of \cite{Poggiolini,Carena,PogECOC,PogArxiv}. If the pre-dispersed signals are Gaussian enough at the end of the first span so as to satisfy the assumptions of the GN model, how come they are less Gaussian further along the system given that the accumulated dispersion increases monotonically? Our own interpretation to this behavior relies once again on the time domain picture of pulse collisions. When the temporal spreading of the launched pulses is larger than the walk-off between channels, all collisions become incomplete, and for the same reasons that we explained earlier the GN model becomes more accurate. Nonetheless, when the system length increases to the extent that the inter-channel walk-off becomes large enough to accommodate full collisions, the deviation from the GN result reappears once again.

When examining the situation in the frequency domain picture, pre-dispersion implies rapid phase variations in the interfering channel's spectrum  (i.e. variations in the phase of $\tilde g(\omega)$ in the notation of \cite{DarOpex}). While the SON coefficient $\chi_1$ is not affected by the spectral phase, the FON coefficient $\chi_2$ reduces considerably in this situation, since the fourth-order correlation terms (Eq. 24 in \cite{DarOpex}) lose coherence. We note however, that since with all relevant modulation formats, the quantity $\frac{\lip |b|^4\rip}{\langle |b|^2\rangle^2} -2$ is negative, the reduction of $\chi_2$ through pre-dispersion always leads to an increase in the NLIN variance and is therefore undesirable.
\begin{figure}
\begin{centering}
\includegraphics[width=0.75\columnwidth]{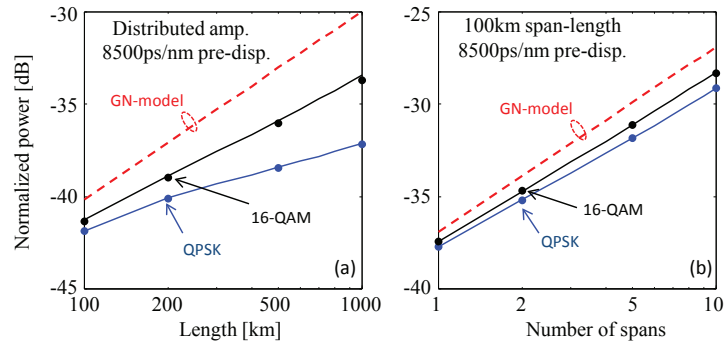}
\caption{The effect of pre-dispersion. Accumulation of the NLIN power (normalized to the received power) with the number of spans. Figures a and b correspond to distributed amplification and span-length of 100 km, respectively, where pre-dispersion of 8500 ps/nm was applied to the injected pulses. The solid lines show the theoretical results given by Eq. (\ref{10}) and the dots represent simulations. The red dashed curve represents the SON, or equivalently the GN model result.}\label{Accum}
\end{centering}
\end{figure}

\section{Conclusions}
We have shown that the previously unexplained dependence of the NLIN variance on pre-dispersion, modulation format and on the number of amplified spans, is accounted for by the FON term, which follows from the correct treatment of the signal's statistics \cite{DarOpex}. Excellent agreement between theory and simulations has been demonstrated in all of our simulations, suggesting that in the range of parameters that we have tested, the inclusion of additional correction terms, which were presented in \cite{PogArxiv} is not necessary.
The relative magnitude of the FON term is largest in single span systems, or in systems using distributed amplification, and it reduces notably in the case of lumped amplification with a large number of spans. Similarly, the relative significance of phase noise (which is included both in the SON and the FON terms) is largest in single span systems, or in systems with distributed amplification, although it remains significant in all the cases that we have tested.


%
\section*{Appendix: Computation of $\chi_1$ and $\chi_2$}
The extraction of the analytical curves in Figs. 1 and 3 --5 relies on the computation of the SON and FON coefficients $\chi_1$ and $\chi_2$, which requires summation over three and five indices, respectively. Multi-dimensional summations are extremely inefficient in brute-force computation, and hence we have adopted the Monte-Carlo integration method \cite{MCI} for evaluating these quantities. We provide a code (written in Matlab) that computes $\chi_1$, $\chi_2$ (using Eqs. (26) and (27) of \cite{DarOpex} including receiver matched filtering that removes products that fall outside of the received channel bandwidth), and allows the evaluation of the implied NLIN variance according to Eq. (\ref{10}). The program also evaluates the relative error in the computed NLIN variance, where in all of the numerical curves presented in this paper the number of integration points $N$ was large enough to ensure that the relative error was well under 1\%.

{For the reader's convenience, in addition to providing the tools for reproducing the curves presented in this paper, we include blocks that compute the variance of nonlinear intra-channel noise, as well as additional inter-channel terms that contribute to NLIN when the guard-band between WDM channels is much smaller than the channel bandwidth, and which were first reported in} \cite{PogArxiv}. {It can be easily verified that the contribution of these terms is negligible with the typical system parameters assumed in this paper, although they may play a role in the case of future densely spaced super-channels. The option of describing a polarization multiplexed link is also included. The program assumes perfect Nyquist pulses and homogeneous fiber spans, but it can be readily extended to an arbitrary pulse-shape and to systems with different span length and fiber dispersions.}

The runtime for the computation of the SON and FON coefficients $\chi_1$ and $\chi_2$ on a standard PC with an i5 processor is of the order of 0.5 seconds, whereas the computation of all (intra and inter-channel) NLIN terms is performed in less than 2 seconds. We note that polarization multiplexing does not affect the run-time of the code (although it more than doubles the runtime of a full split-step simulation).

\begin{lstlisting}

function main()
%% System parameters
clear;clc;tic;
PolMux = 0;     % 0 when single polarization, 1 with polarization mulltiplexing
gamma = 1.3;    % Nonlinearity coefficient in [1/W/km]
beta2 = 21;     % Dispersion coefficient [ps^2/km]
alpha = 0.2;    % Fiber loss coefficient [dB/km]
Nspan = 5;      % Number of spans
L = 100;        % Span length [km]
PD = 0;         % Pre-dispersion [ps^2]
PdBm = -2;      % Average input power [dBm]
BaudRate = 32;  % Baud-rate [GHz]
ChSpacing = 50; % Channel spacing [GHz]
kur = 1.32;     % Second order modulation factor <|a|^4>/<|a|^2>^2
kur3 = 1.96;    % Third order modulation factor <|a|^6>/<|a|^2>^3
N = 1000000;    % Number of integration points in algorithm [14]. Should 
                % be set such that the relative error is desirably small. 

%% Monte-Carlo integration
alpha_norm = 0.2/10*log(10); 
T=1000/BaudRate; 
P0 = 10^((PdBm-30)/10);
beta2_norm = beta2/T^2;
PD_norm = PD/T^2;
ChSpacing_norm = ChSpacing./BaudRate;

%% calculate inter-channel nonlinear noise variance according to Eq. (1)
[NLIN_var chi1 chi2 Err] = calc_interChannel(gamma,beta2_norm,alpha_norm,...
                            Nspan,L,PD_norm,P0,kur,kur3,N,PolMux,ChSpacing_norm);

%% display
disp('%%%%%%%%%%%%%%%%%%%%%%%%%%%%%%%%%%%%%%%%%%%%%%');
if(PolMux == 1)
    disp('%%%Polarization Multiplexed case is considered%%%');
end
disp('%%Results correspond to a single interferer%%%')
disp(['(1) chi_1 = ' num2str(chi1) ', chi_2 = ' num2str(chi2)]);
disp(['(2) NLIN variance according to Eq. (1) is ' num2str(NLIN_var)...
        ' Watts (' num2str(10*log10(NLIN_var*1000)) 'dBm).'...
        ' Relative computation error is ', num2str(Err*100),'%']);
    
%% calculate the contribution of the additional inter-channel terms of [7]
NLIN_var_addTerm = calc_interChannel_addTerms(gamma,beta2_norm,alpha_norm,...
                            Nspan,L,PD_norm,P0,kur,kur3,N,PolMux,ChSpacing_norm);

%% calculate intra-channel nonlinear noise 
%% NLIN_var_intra(1) is the intra-channel nonlinear noise variance 
%% NLIN_var_intra(2) is due to the nonlinear broadening of the adjacent interferer 
NLIN_var_intra = calc_intraChannel(gamma,beta2_norm,alpha_norm,...
                            Nspan,L,PD_norm,P0,kur,kur3,N,PolMux,ChSpacing_norm);
NLIN_var_addTerm = NLIN_var_addTerm + NLIN_var_intra(2);

%%Continue display
disp(['(3) Contribution of additional inter-channel interference '...
        'terms of [7] is ' num2str(NLIN_var_addTerm) ' Watts ('...
        num2str(10*log10(NLIN_var_addTerm*1000)) 'dBm)']);
disp(['(4) Total NLIN variance (2)+(3) is ' num2str(NLIN_var+NLIN_var_addTerm)... 
        ' Watts (' num2str(10*log10((NLIN_var+NLIN_var_addTerm)*1000)) 'dBm)']);
disp(['(5) Intra-Channel nonlinear noise variance is ' num2str(NLIN_var_intra(1))...
        ' Watts (' num2str(10*log10(NLIN_var_intra(1)*1000)) 'dBm)']);
toc;
disp('%%%%%%%%%%%%%%%%%%%%%%%%%%%%%%%%%%%%%%%%%%%%%%');
end

%%%%%%%%%%%%%%%%%%%%%%%%%
function [NLIN_var chi1 chi2 Err] = ...
    calc_interChannel(gamma,beta2,alpha,Nspan,L,PD,P0,kur,kur3,N,PolMux,q)

R = 2*pi*(rand(4, N)-0.5*ones(4, N));
Volume = (2*pi)^4;

%% calculate chi1 
w0 = R(1,:)-R(2,:)+R(3,:);
arg1 = (R(2,:)-R(3,:)).*(R(2,:)+2*pi*q-R(1,:));
argPD1 = arg1;
ss1 = exp(1i*argPD1*PD).*(exp(1i*beta2*arg1*L-alpha*L)-1)...
    ./(1i*beta2*arg1-alpha).*(w0<pi).*(w0>-pi);
s1 = abs(ss1.*(1-exp(1i*Nspan*arg1*beta2*L))...
    ./(1-exp(1i*arg1*beta2*L))).^2/Volume;
avgF1 = sum(s1)/N;
chi1 = avgF1*Volume*(4*gamma^2*P0^3);

%% calculate chi2
w3p = -R(2,:)+R(4,:)+R(3,:)+2*pi*q; 
arg2 = (R(2,:)-R(3,:)).*(R(4,:)-R(1,:)+2*pi*q);
argPD2 = arg2;
ss2 = exp(-1i*argPD2*PD).*(exp(-1i*beta2*arg2*L-alpha*L)-1)...
    ./(-1i*beta2*arg2-alpha).*(w3p>-pi+2*pi*q).*(w3p<pi+2*pi*q);
s2 = (1-exp(1i*Nspan*arg1*beta2*L))./(1-exp(1i*arg1*beta2*L)).*ss1...
    .*(1-exp(-1i*Nspan*arg2*beta2*L))...
    ./(1-exp(-1i*arg2*beta2*L)).*ss2/Volume;
avgF2 = real(sum(s2))/N;
chi2 = avgF2*Volume*(4*gamma^2*P0^3);

%% calculate NLIN
NLIN_var = chi1+(kur-2)*chi2;
if(PolMux == 1)
    NLIN_var = (9/8)^2*16/81*(NLIN_var+2*chi1/4+(kur-2)*chi2/4);
end

%% calculate the root mean square relative error
if(PolMux == 0)
    Err = (sum((s1-avgF1+(kur-2)*(real(s2)-avgF2)).^2)...
        /(N-1))^.4/(avgF1+(kur-2)*avgF2)/sqrt(N);
else
    Err = (sum(((9/8)^2*16/81*(6/4*(s1-avgF1)+5/4*(kur-2)*...
        (real(s2)-avgF2))).^2)/(N-1))^.4/((9/8)^2*16/81*...
        (6/4*avgF1+5/4*(kur-2)*avgF2))/sqrt(N);
end

end

%%%%%%%%%%%%%%%%%%%%%%%%%
function [NLIN_var] = calc_interChannel_addTerms(gamma,beta2,alpha,...
                                Nspan,L,PD,P0,kur,kur3,N,PolMux,q)

R = 2*pi*(rand(4, N)-0.5*ones(4, N));

%% calculate X21
w0 = R(1,:)-R(2,:)+R(3,:)+2*pi*q;
arg1 = (R(2,:)-R(3,:)-2*pi*q).*(R(2,:)-R(1,:));
argPD1 = arg1;
ss1 = exp(1i*argPD1*PD).*(exp(1i*beta2*arg1*L-alpha*L)-1)...
    ./(1i*beta2*arg1-alpha).*(w0<pi).*(w0>-pi);
s1 = abs(ss1.*(1-exp(1i*Nspan*arg1*beta2*L))./(1-exp(1i*arg1*beta2*L))).^2;
X21 = sum(s1)*(gamma^2*P0^3)/N;

%% calculate X22
w1 = R(1,:)-R(2,:)+R(4,:);
arg2 = (w1-R(3,:)-2*pi*q).*(R(2,:)-R(1,:));
argPD2 = arg2;
ss2 = exp(-1i*argPD2*PD).*(exp(-1i*beta2*arg2*L-alpha*L)-1)...
    ./(-1i*beta2*arg2-alpha).*(w1<pi).*(w1>-pi);
s2 = (1-exp(1i*Nspan*arg1*beta2*L))./(1-exp(1i*arg1*beta2*L)).*ss1...
    .*(1-exp(-1i*Nspan*arg2*beta2*L))./(1-exp(-1i*arg2*beta2*L)).*ss2;
X22 = real(sum(s2))*(gamma^2*P0^3)/N;

%% calculate X23
w2 = R(1,:)+R(2,:)-R(3,:)-2*pi*q;
arg1 = (R(3,:)+2*pi*q-R(2,:)).*(R(3,:)+2*pi*q-R(1,:));
argPD1 = arg1;
ss3 = exp(1i*argPD1*PD).*(exp(1i*beta2*arg1*L-alpha*L)-1)...
    ./(1i*beta2*arg1-alpha).*(w2<pi).*(w2>-pi);
s3 = abs(ss3.*(1-exp(1i*Nspan*arg1*beta2*L))./(1-exp(1i*arg1*beta2*L))).^2;
X23 = sum(s3)*(gamma^2*P0^3)/N;

%% calculate X24
w3 = R(1,:)-R(4,:)+R(2,:); 
arg2 = (R(3,:)+2*pi*q-R(4,:)).*(R(3,:)+2*pi*q-w3);
argPD2 = arg2;
ss4 = exp(-1i*argPD2*PD).*(exp(-1i*beta2*arg2*L-alpha*L)-1)...
    ./(-1i*beta2*arg2-alpha).*(w3<pi).*(w3>-pi);
s4 = (1-exp(1i*Nspan*arg1*beta2*L))./(1-exp(1i*arg1*beta2*L)).*ss3...
    .*(1-exp(-1i*Nspan*arg2*beta2*L))./(1-exp(-1i*arg2*beta2*L)).*ss4;
X24 = real(sum(s4))*(gamma^2*P0^3)/N;

%% calculate NLIN
NLIN_var = 4*X21+4*(kur-2)*X22+2*X23+(kur-2)*X24;
if(PolMux == 1)
    NLIN_var = (9/8)^2*16/81*(NLIN_var+2*X21+(kur-2)*X22+X23+0*(kur-2)*X24);
end

end

%%%%%%%%%%%%%%%%%%%%%%%%%
function [NLIN_var] = ...
    calc_intraChannel(gamma,beta2,alpha,Nspan,L,PD,P0,kur,kur3,N,PolMux,q)

if(exist('q')==0) q = 0; end;
R = 2*pi*(rand(5, N)-0.5*ones(5, N));

%% calculate X1
w0 = R(1,:)-R(2,:)+R(3,:);
argInB = (w0<pi).*(w0>-pi);
argOutB = (w0<pi+2*pi*q).*(w0>-pi+2*pi*q);
arg1 = (R(2,:)-R(3,:)).*(R(2,:)-R(1,:));
argPD1 = arg1;
ss1 = exp(1i*argPD1*PD).*(exp(1i*beta2*arg1*L-alpha*L)-1)...
    ./(1i*beta2*arg1-alpha);
s1 = abs(ss1.*(1-exp(1i*Nspan*arg1*beta2*L))./(1-exp(1i*arg1*beta2*L))).^2;
X1 = [sum(s1.*argInB) sum(s1.*argOutB)]*(gamma^2*P0^3)./N;

%% calculate X0
s0 = ss1.*(1-exp(1i*Nspan*arg1*beta2*L))./(1-exp(1i*arg1*beta2*L));
X0 = [abs(sum(s0.*argInB)/N).^2 abs(sum(s0.*argOutB)/N).^2]*(gamma^2*P0^3);

%% calculate X2
w1 = -R(2,:)+R(4,:)+R(3,:); 
arg2 = (R(2,:)-R(3,:)).*(R(4,:)-R(1,:));
argPD2 = arg2;
ss2 = exp(-1i*argPD2*PD).*(exp(-1i*beta2*arg2*L-alpha*L)-1)...
    ./(-1i*beta2*arg2-alpha).*(w1<pi).*(w1>-pi);
s2 = (1-exp(1i*Nspan*arg1*beta2*L))./(1-exp(1i*arg1*beta2*L)).*ss1...
    .*(1-exp(-1i*Nspan*arg2*beta2*L))./(1-exp(-1i*arg2*beta2*L)).*ss2;
X2 = [real(sum(s2.*argInB)) real(sum(s2.*argOutB))]*(gamma^2*P0^3)./N;

%% calculate X21
w2 = R(4,:)-R(1,:)-R(3,:); 
arg2 = (R(2,:)-R(4,:)).*(R(2,:)-w2);
argPD2 = arg2;
ss2 = exp(-1i*argPD2*PD).*(exp(-1i*beta2*arg2*L-alpha*L)-1)...
    ./(-1i*beta2*arg2-alpha).*(w2<pi).*(w2>-pi);
s21 = (1-exp(1i*Nspan*arg1*beta2*L))./(1-exp(1i*arg1*beta2*L)).*ss1...
    .*(1-exp(-1i*Nspan*arg2*beta2*L))./(1-exp(-1i*arg2*beta2*L)).*ss2;
X21 = [real(sum(s21.*argInB)) real(sum(s21.*argOutB))]*(gamma^2*P0^3)./N;

%% calculate X3
w3 = R(1,:)-R(2,:)+R(4,:)+R(3,:)-R(5,:);
arg3 = (R(4,:)-R(5,:)).*(R(4,:)-w3);
argPD3 = arg3;
ss3 = exp(-1i*argPD3*PD).*(exp(-1i*beta2*arg3*L-alpha*L)-1)...
    ./(-1i*beta2*arg3-alpha).*(w3<pi).*(w3>-pi);
s3 = (1-exp(1i*Nspan*arg1*beta2*L))./(1-exp(1i*arg1*beta2*L)).*ss1...
    .*(1-exp(-1i*Nspan*arg3*beta2*L))./(1-exp(-1i*arg3*beta2*L)).*ss3;
X3 = [real(sum(s3.*argInB)) real(sum(s3.*argOutB))]*(gamma^2*P0^3)./N;

%% calculate NLIN
NLIN_var = 2*X1+(kur-2)*(4*X2+X21)+(kur3-9*kur+12)*X3-(kur-2)^2*X0;
if(PolMux == 1)
   NLIN_var = (9/8)^2*16/81*(NLIN_var+X1+(kur-2)*X2);
end

end
\end{lstlisting}

\section*{Acknowledgment}
The authors would like to acknowledge financial support from the Israel Science Foundation (grant 737/12). Ronen Dar would like to acknowledge the support of the Adams Fellowship of the Israel Academy of Sciences and Humanities, the Yitzhak and Chaya Weinstein Research Institute, and the Feder Family Award.

\end{document}